\documentclass[aps, prb, reprint, preprintnumbers, superscriptaddress, showpacs, floatfix]{revtex4-1}

\usepackage{graphicx}
\usepackage{dcolumn}

\begin{document}

\preprint{Ver. 3}

\title{Two distinct superconducting states in KFe$_2$As$_2$ under high pressure}


\author{Taichi Terashima}
\affiliation{National Institute for Materials Science, Tsukuba, Ibaraki 305-0003, Japan}
\author{Kunihiro Kihou}
\affiliation{National Institute of Advanced Industrial Science and Technology (AIST), Tsukuba, Ibaraki 305-8568, Japan}
\author{Kaori Sugii}
\author{Naoki Kikugawa}
\author{Takehiko Matsumoto}
\affiliation{National Institute for Materials Science, Tsukuba, Ibaraki 305-0003, Japan}
\author{Shigeyuki Ishida}
\author{Chul-Ho Lee}
\author{Akira Iyo}
\author{Hiroshi Eisaki}
\affiliation{National Institute of Advanced Industrial Science and Technology (AIST), Tsukuba, Ibaraki 305-8568, Japan}
\author{Shinya Uji}
\affiliation{National Institute for Materials Science, Tsukuba, Ibaraki 305-0003, Japan}


\date{\today}

\begin{abstract}
We report measurements of ac magnetic susceptibility $\chi_{ac}$ and de Haas-van Alphen (dHvA) oscillations in KFe$_2$As$_2$ under high pressure up to 24.7 kbar.
The pressure dependence of the superconducting transition temperature $T_c$ changes from negative to positive across $P_c \sim 18$ kbar as previously reported.
The ratio of the upper critical field to $T_c$, i.e, $B_{c2} / T_c$, is enhanced above $P_c$, and the shape of $\chi_{ac}$ vs field curves qualitatively changes across $P_c$.
DHvA oscillations smoothly evolve across $P_c$, indicating no drastic change in the Fermi surface up to 24.7 kbar.
Three dimensionality increases with pressure, while effective masses show decreasing trends.
We suggest a crossover from a nodal to a full-gap $s$ wave as a possible explanation. 
\end{abstract}


\maketitle



\newcommand{\ud}{\mathrm{d}}
\def\degree{\kern-.2em\r{}\kern-.3em}

\section{Introduction}

\begin{figure}[!h]
\includegraphics[width=8.4cm]{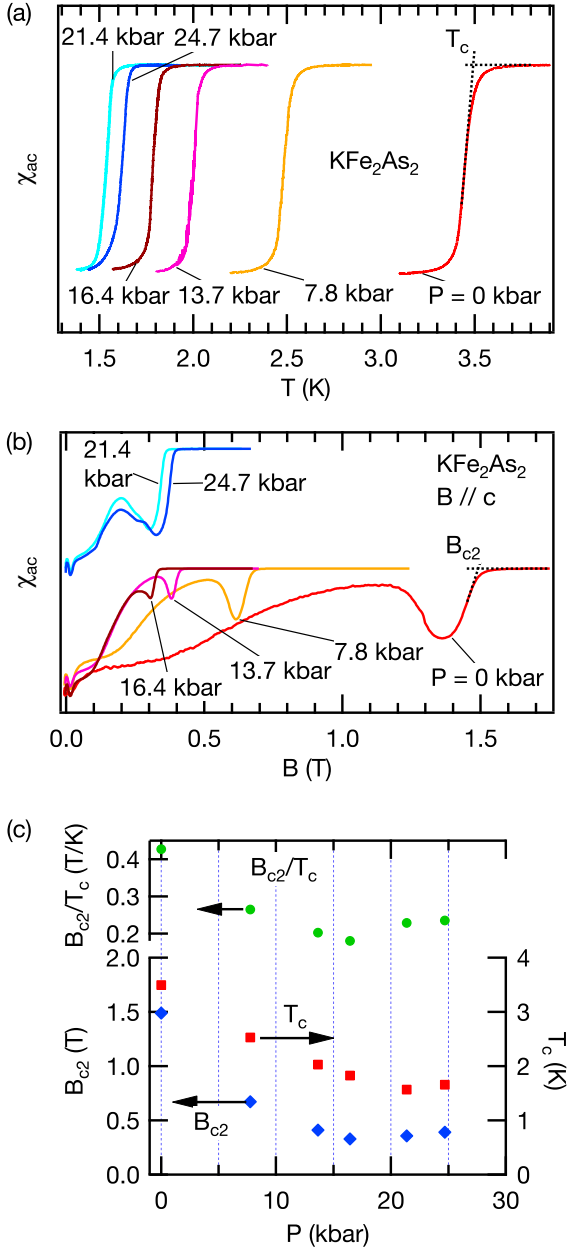}
\caption{\label{fig1}(color online).  (a) AC magnetic susceptibility $\chi_{ac}$ as a function of $T$ at different pressures.  (b) $\chi_{ac}$ as a function of the field $B$ applied parallel to the $c$ axis at different pressures.  The curves for $P$ = 21.4 and 24.7 kbar are shifted vertically.  The measurement temperature is 0.02$T_c$ or lower at each pressure.  (c) Pressure dependencies of $T_c$, $B_{c2}$, and $B_{c2}/T_c$.  The definitions of $T_c$ and $B_{c2}$ are shown in (a) and (b), respectively.}   
\end{figure}

Since the discovery of superconductivity at $T_c$ = 26 K in LaFeAs(O, F) by Kamihara \textit{et al}.,\cite{Kamihara08JACS} iron-based high-$T_c$ superconductivity has been one of the focuses of research in the condensed-matter physics community. 
One of notable features of the iron-based superconductors is that, unlike the cuprates, their gap structures are not universal: while some are fully gapped, others have nodal gap structures.\cite{[{See. for a review, }] Hirschfeld11RPP}
Especially intriguing is the case of (Ba$_{1-x}$K$_x$)Fe$_2$As$_2$: thermodynamic and other measurements indicate that the gap structure changes from a fully gapped one near the optimal doping ($x \sim$ 0.4)\cite{Hashimoto09PRL, Mu09PRB, Popovich10PRL, Luo09PRB, Yashima09JPSJ} to a nodal one at $x$ = 1, i.e, KFe$_2$As$_2$.\cite{Fukazawa09JPSJ_KFA, Hashimoto10PRB, Dong10PRL, Reid12PRL, Okazaki12Science, Kittaka14JPSJ}
A laser ARPES (angle-resolved photoemission spectroscopy) study on KFe$_2$As$_2$ has found an $s$-wave gap with accidental nodes on one of the $\Gamma$-centered Fermi surface cylinders,\cite{Okazaki12Science} although there is a claim of a $d$-wave gap based on thermal conductivity measurements.\cite{Reid12PRL} 
Theoretical studies based on spin-fluctuations approaches indicate that various gap structures with $s$- or $d$-wave symmetries are in close competition.\cite{Thomale11PRL, Maiti11PRL, Suzuki11PRB, Maiti12PRB}

Recently, Tafti \textit{et al}. have found from resistivity measurements that the pressure dependence of $T_c$ in KFe$_2$As$_2$ changes from negative to positive at a critical pressure $P_c = 17.5$ kbar.\cite{Tafti13NatPhys}
Since the resistivity and Hall coefficient vary smoothly across $P_c$, it has been claimed that this change is not due to a change in the Fermi surface but due to a change of pairing symmetry.
In this paper, our ac magnetic susceptibility $\chi_{ac}$ measurements confirm that the pressure dependence of bulk $T_c$ in KFe$_2$As$_2$ does change at $P_c$.
We show that the ratio $B_{c2}/T_c$, where $B_{c2}$ is the upper critical field for $B \parallel c$, is enhanced above $P_c$ and that magnetic responses in the superconducting states below and above $P_c$ are markedly different.
We also show that de Haas-van Alphen (dHvA) oscillations vary smoothly across $P_c$, suggesting absence of Fermi surface (FS) reconstruction at $P_c$ in line with Ref.~\onlinecite{Tafti13NatPhys}. 
We discuss implications of these observations. 

\section{Experiment}

High-quality single crystals of KFe$_2$As$_2$ were prepared by a self-flux method.\cite{Kihou10JPSJ}
A sample placed in a balanced pick-up coil was pressurized in a NiCrAl piston-cylinder pressure cell, which was loaded in a $^3$He-$^4$He dilution refrigerator equipped with a 20 T superconducting magnet.
The pressure transmitting media was Daphne 7474, which remains liquid up to 37 kbar at room temperature and ensures highly hydrostatic pressure generation.\cite{Murata08RSI} 
A Manganin wire gauge was used to determine the pressure at low temperatures.\cite{Terashima09JPSJ_EFA_erratum}

\section{Results and discussion}

Figure 1(a) shows $\chi_{ac}$ as a function of temperature $T$ for pressures up to 24.7 kbar.
The superconducting transition shits to lower temperatures up to 16.4 kbar ($< P_c$) but moves to higher temperatures as the pressure is increased from 21.4 ($> P_c$) to 24.7 kbar.
The size of the diamagnetic signal does not change with $P$, indicating that the superconducting volume fraction does not change with $P$.
Figure 1(b) shows $\chi_{ac}$ as a function of magnetic field $B$ for pressures up to 24.7 kbar.
The upper critical field $B_{c2}$ decreases with $P$ for $P < P_c$ but increases for $P > P_c$ similarly to $T_c$.
Figure 1(c) shows $T_c$, $B_{c2}$, and $B_{c2}/T_c$ as a function of pressure.
\footnote{The ratio $B_{c2}/T_c$ would be related to the effective mass in the case of single-band superconductors without the Pauli paramagnetic effect, but its interpretation in multiband superconductors like KFe$_2$As$_2$ is not straightforward.}
Although the present values of $T_c$ for $P > P_c$ are slightly lower than those reported in Ref.~\onlinecite{Tafti13NatPhys},  the variation of $T_c$ qualitatively agrees with Ref.~\onlinecite{Tafti13NatPhys} and also with a most recent article.\cite{Taufour14condmat}
Low-pressure data are also consistent with Ref.~\onlinecite{Bud'ko12PRB}.
A new observation here is that the ratio $B_{c2}/T_c$ grows across $P_c$: it decreases with $P$ up to 16.4 kbar but takes a larger value at 21.4 kbar.
We also notice that the $\chi_{ac}(B)$ curves change qualitatively across $P_c$ [Fig. 1(b)].
At $P = 0$ kbar, $\chi_{ac}$ shows a rounded downward peak just below $B_{c2}$.
This is due to the peak effect.
\cite{[{The critical current density in a superconductor sometimes shows an anomalous peak just below $B_{c2}$, indicating an enhancement of pinning force acting on vortices.
This is called peak effect.
Although its origin is not completely understood yet, it is broadly associated with an order-disorder transition of a vortex lattice.
See, for example, }] Giamarchi01condmat}
Figure 1(b) shows that the peak effect becomes smaller and smaller as the pressure is increased up to 16.4 kbar.
However, as the pressure is further increased to 21.4 kbar ($> P_c$), the peak effect strengthens abruptly.
The shape of the peak is also different: a shoulder appears on the lower field side of the peak above $P_c$.
The clear difference of response of the vortex lattice to the applied field between below and above $P_c$ as well as the increase in $B_{c2}/T_c$ above $P_c$ indicates that there are two distinct superconducting states below and above $P_c$.

\begin{figure}
\includegraphics[width=4cm]{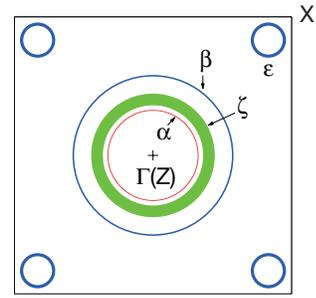}
\caption{\label{fig2}(color online).  Fermi surface cross-sections observed via dHvA measurements at ambient pressure.  The in-plane anisotropy is neglected.  The line thickness indicates the magnitude of the $c$-axis dispersion.  Reproduced from Ref.~\onlinecite{Terashima13PRBdHvA}}   
\end{figure}

Before describing dHvA data under high pressure, we review the Fermi surface in KFe$_2$As$_2$ at ambient pressure.
As schematically shown in Fig. 2, the Fermi surface consists of three hole cylinders $\alpha$, $\zeta$, and $\beta$ at the $\Gamma$ point of the Brillouin zone (BZ) and small hole cylinders $\epsilon$ near the X point.\cite{Sato09PRL, Terashima10JPSJ, Yoshida11JPCS, Terashima13PRBdHvA}
Each cylinder has the minimum and maximum orbits giving rise to two dHvA frequencies labeled with subscripts $l$ and $h$ hereafter, except that only one frequency has been found for the $\beta$ cylinder.\cite{Terashima13PRBdHvA}
The $\alpha_l$ and $\zeta_l$ orbits are close enough for magnetic breakdown to occur at certain points of the orbits.
See the $P$ = 0 kbar spectrum in Fig. 4(a): small peaks except $\alpha_h$ between the $\alpha_l$ and $\zeta_l$ peaks are magnetic breakdown frequencies and their amplitudes are quickly suppressed as field is decreased.\cite{Terashima13PRBdHvA}

\begin{figure}
\includegraphics[width=8.4cm]{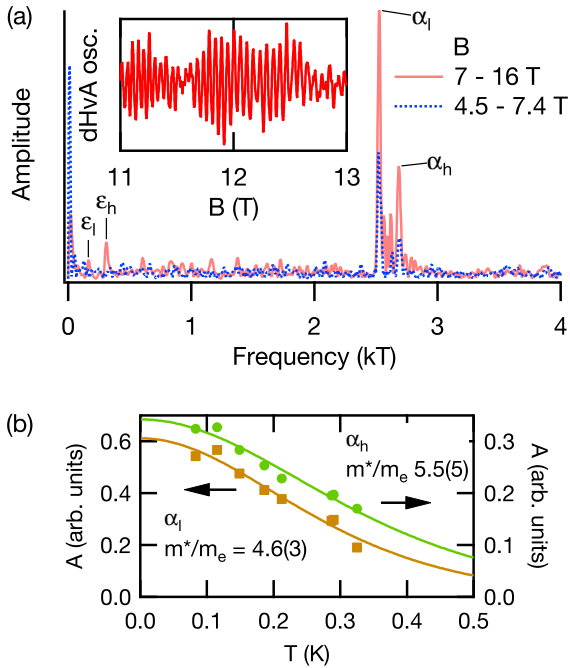}
\caption{\label{fig3}(color online).  (a) Fourier transforms of dHvA oscillations at $P$ = 21.4 kbar in two different field windows.  The two large peaks in the low-field spectrum (dotted) are identified as the $\alpha_{l, h}$ fundamental frequencies.  The inset shows part of the recorded oscillations.  (b) Temperature dependences of the amplitudes of the $\alpha_{l, h}$ frequencies.  Solid lines are fits to the Lifshitz-Kosevich formula, from which the effective masses are estimated as indicated.}   
\end{figure}

DHvA oscillations have been observed at all the measured pressures.
For example, Fig. 3(a) shows Fourier transforms of dHvA oscillations in two field windows measured at $P$ = 21.4 kbar.
Part of the oscillations is shown in the inset. 
Two frequency peaks below $F \lesssim 0.3$ kT are identified as the $\epsilon_l$ and $\epsilon_h$ frequencies.
Among many peaks appearing in a frequency region between 2.5 and 3 kT, two peaks that retain substantial amplitudes in the low-field spectrum (dotted line) are identified as the $\alpha_l$ and $\alpha_h$ fundamental frequencies.
The rest can be ascribed to magnetic breakdown frequencies between $\alpha_l$ and $\zeta_l$, although it is difficult to identify the latter.
Figure 3(b) shows the temperature dependences of the amplitudes of $\alpha_l$ and $\alpha_h$.
By fitting the Lifshitz-Kosevich formula \cite{Shoenberg84} to them, we obtain effective masses of 4.6(3) and 5.5(5) $m_e$, respectively, where $m_e$ is the free electron mass.

\begin{figure}
\includegraphics[width=8cm]{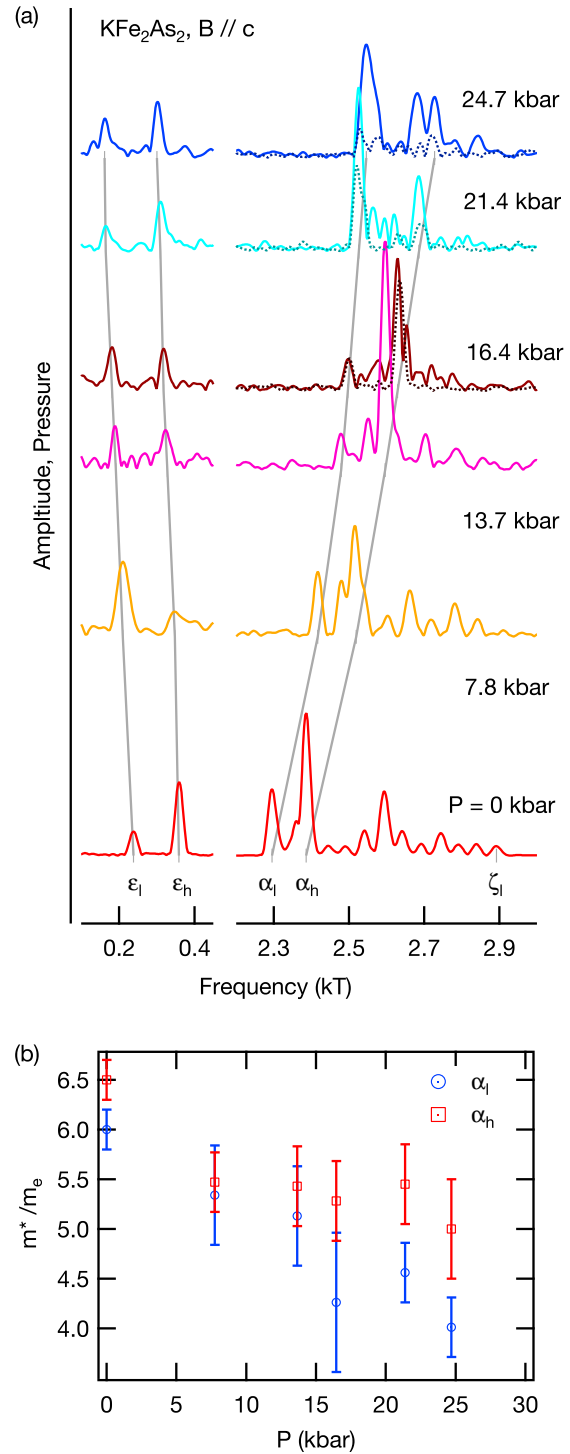}
\caption{\label{fig4}(color online).  (a) Fourier transforms of dHvA oscillations at different pressures.  For $P$ = 16.4, 21.4, and 24.7 kbar, high-field (solid) and low-field (dotted) transforms are shown as in Fig. 3(a) to identify the $\alpha_{l, h}$ fundamental frequencies.  (b) Pressure dependencies of the effective masses associated with the $\alpha_{l, h}$ fundamental frequencies.  In both (a) and (b), the data for $P$ = 0 kbar are taken from Ref.~\onlinecite{Terashima13PRBdHvA}.}   
\end{figure}

Figure 4(a) shows the Fourier spectra for all the measured pressures.
The $\epsilon_{l, h}$ frequencies decrease with pressure, while the $\alpha_{l, h}$ frequencies increase.
The frequencies of $\epsilon_l$, $\epsilon_h$, $\alpha_l$, and $\alpha_h$ at $P$ = 24.7 kbar are 68, 84, 111, and 114\% of those at $P$ = 0 kbar.
Since the lattice is compressed by pressure, the Brillouin zone expands with pressure and hence dHvA frequencies would increase even without change in the electronic structure.
Although there is no high-pressure structural data on KFe$_2$As$_2$, a neutron study on BaFe$_2$As$_2$ shows that the $a$ axis shrinks by about 1\% at $P$ = 25 kbar, \cite{Kimber09NatMat} corresponding to 2 \% expansion of the Brillouin zone.
The observed changes are larger than this and are opposite in the cases of the $\epsilon_{l, h}$ frequencies.

We first consider the possibility of a Lifshitz transition at $P_c$.
It seems that the $\Gamma$-centered cylinders are too big to disappear and too small to touch the Brillouin zone boundary by the application of a moderate pressure of 25 kbar (see Fig. 2), and indeed the $\alpha$ frequencies continue to exist above $P_c$.
The only possibility to be seriously considered is disappearance or fragmentation of the $\epsilon$ cylinder.
However, Fig. 4(a) clearly shows that the $\epsilon$ cylinder continues to exist above $P_c$.
We therefore conclude that no Lifshitz transition occurs at $P_c$ in line with Ref.~\onlinecite{Tafti13NatPhys}.
The pressure variation of the $\epsilon_{l, h}$ frequencies is smooth across $P_c$.
If we extrapolate it to higher pressures, we can expect that the cylinder will be divided at the position of the minimum cross-section, i.e., $F_{\epsilon_l} \to 0$, near 70 kbar.
It would be interesting to see what happens there.

We now turn to the global structure of the Fermi surface above $P_c$.
At ambient pressure, the $\epsilon$, $\alpha$, $\zeta$, and $\beta$ cylinders have been estimated from the dHvA frequencies to occupy 1.1$\times$4, 8.4, 13.0, and 25.6\% of the Brillouin zone (\% BZ) respectively (there are four $\epsilon$ cylinders per the Brillouin zone). \cite{Terashima13PRBdHvA}
The present dHvA data at $P$ = 24.7 kbar indicate that the $\epsilon$ and $\alpha$ cylinders occupy 0.8$\times$4 and 9.2\% BZ, respectively, where we have used the average of the minimum and maximum frequencies and have taken into account the 2\% expansion of the Brillouin zone cross-section.
As explained above, magnetic breakdown frequencies between $\alpha_l$ and $\zeta_l$ are observed above $P_c$ [Fig. 4(a)].
It indicates that the sizes of the two cylinders stay close above $P_c$, and hence we can assume that the $\zeta$ cylinder grows by about 1\% BZ from $P$ = 0 to 24.7 kbar as $\alpha$ does.
To conserve the total Fermi surface volume, the $\beta$ cylinder would shrink by about 1\% BZ at $P$ = 24.7 kbar.
These estimates indicate that the global structure of the Fermi surface hardly changes up to $P$ = 24.7 kbar ($> P_c$).

Nonetheless, there are noteworthy changes in the electronic structure.
Firstly, the three-dimensionality grows with pressure as expected.
To quantify it, $\Delta F / F_{av}$ increases from 0.40 to 0.59 for the $\epsilon$ cylinder and from 0.038 to 0.069 for the $\alpha$ cylinder as the pressure is increased up to $P$ = 24.7 kbar, where $\Delta F$ and $F_{av}$ are the difference and the average of the minimum and maximum frequencies.
Secondly, there is a decreasing trend in effective masses.
Fig. 4(b) shows the effective masses of $\alpha_{l, h}$ as a function of pressure.
Although error bars are large, a general decreasing trend can be seen both below and above $P_c$.
The effective mass $m^*$ can be expressed as $m^* = m_{band}(1+\lambda)$, $m_{band}$ being the band mass.
The mass enhancement $(1+\lambda)$ is due to interactions between electrons and bosons such as phonons and spin fluctuations.
Since changes in the sizes of the Fermi surface cylinders are not appreciable up to $P$ = 24.7 kbar as noted above, we may assume $m_{band}$ to be constant.
Then, the decreasing trend in $m^*$ translates into that in $\lambda$.

We now discuss implications of our results.
The increasing three-dimensionality and decreasing $\lambda$ are generally considered unfavorable to superconductivity, and hence seem at odds with the increasing $T_c$ above $P_c$.
Let us consider a McMillan-type formula $T_c \sim T_o \exp[-(1+\lambda) / \lambda]$.
The characteristic boson energy $T_o$ here is the spin-fluctuation energy $T_{SF}$, which is roughly inversely proportional to $\gamma$ and hence to $(1+\lambda)$, where $\gamma$ is the electronic specific heat coefficient. \cite{Nakamura96JPSJ}
The above formula then produces a broad maximum of $T_c$ as the coupling strength $\lambda$ is varied but not a minimum as observed.
Even with more elaborated treatments of spin-fluctuation mediated superconductivity such as Refs.~\onlinecite{Nakamura96JPSJ, Monthoux01PRB}, it would be difficult to reconcile the minimum of $T_c$ at $P_c$ with the decreasing $\lambda$ within a single-band picture.
The reversal of the pressure dependence of $T_c$ is therefore related to the multiband nature of the superconductivity in KFe$_2$As$_2$.
Namely, it is caused by competition among different intraband and interband pair scatterings.

Finally, we consider whether $P_c$ is a phase transition or crossover ?
The $T_c(P)$ curve reported in Ref.~\onlinecite{Tafti13NatPhys} has a sharp V-shaped minimum at $P_c$.
Since the authors take the view of a $d$-wave symmetry at ambient pressure, they have suggested a phase transition from a $d$ wave to an $s$ wave at $P_c$.
In comparison, the anomaly at $P_c$ in our $T_c(P)$ data appears rather weak and gradual.
Also, the laser ARPES study has provided strong evidence that the gap structure at ambient pressure is an $s$-wave one with accidental nodes, and results of a magnetic-field-angle dependent specific heat study are compatible with it. \cite{Okazaki12Science, Kittaka14JPSJ}
We therefore suggest another possibility that the minimum of $T_c$ at $P_c$ may be due to a crossover from a nodal $s$ wave to a full-gap $s$ wave.
Since the symmetry remains an $s$ wave, it is not a phase transition. \cite{Hirschfeld11RPP}
Existence of nodes on the Fermi surface is basically unfavorable to $T_c$ because part of strong pair scattering contributes destructively to pairing. \cite{[{See, for example, }] Kuroki01PRB}
Thus removal of nodes may result in enhancement of $T_c$.
We here pay spacial attention to a recent theoretical work. \cite{Maiti12PRB}
It successfully predicts a gap structure that is compatible with the laser ARPES data, using interaction parameters estimated from band structure calculations.
The predicted gap structure is an $s$-wave one in which the gap function basically changes sign between the two $\Gamma$-centered hole pockets in the unfolded zone, which correspond to our $\alpha$ and $\zeta$ cylinders, and has nodes on one of the pockets corresponding to our $\zeta$ cylinder.
More importantly, it also shows that if the intraband scattering within the $\alpha$ cylinder is only slightly weakened, the nodes are readily removed.

\section{Summary}

We have confirmed from bulk $T_c$ measurements that the pressure dependence of $T_c$ in KFe$_2$As$_2$ changes sign at $P_c$.
The ratio $B_{c2}/T_c$ is enhanced above $P_c$, and magnetic responses of the vortex lattice show qualitative changes across $P_c$.
These suggest two different superconducting states below and above $P_c$.
On the other hand, our dHvA data have shown no drastic change across $P_c$.
The global structure of the Fermi surface hardly changes up to the highest pressure of 24.7 kbar, and no Lifshitz transition occurs at $P_c$.
Only changes that we have observed are the increasing three-dimensionality and the decreasing trend in the effective mass.
They are generally unfavorable to superconductivity within a single-band picture.
We have argued that the reversal of the pressure dependence of $T_c$ is related to competition between different pair scattering processes and have proposed that it may be explained by crossover from a nodal to a full-gap $s$ wave superconductivity.

\begin{acknowledgments}
We thank Hiroaki Ikeda and Kazuhiko Kuroki for valuable discussions.
\end{acknowledgments}


\begin{thebibliography}{34}%
\makeatletter
\providecommand \@ifxundefined [1]{%
 \@ifx{#1\undefined}
}%
\providecommand \@ifnum [1]{%
 \ifnum #1\expandafter \@firstoftwo
 \else \expandafter \@secondoftwo
 \fi
}%
\providecommand \@ifx [1]{%
 \ifx #1\expandafter \@firstoftwo
 \else \expandafter \@secondoftwo
 \fi
}%
\providecommand \natexlab [1]{#1}%
\providecommand \enquote  [1]{``#1''}%
\providecommand \bibnamefont  [1]{#1}%
\providecommand \bibfnamefont [1]{#1}%
\providecommand \citenamefont [1]{#1}%
\providecommand \href@noop [0]{\@secondoftwo}%
\providecommand \href [0]{\begingroup \@sanitize@url \@href}%
\providecommand \@href[1]{\@@startlink{#1}\@@href}%
\providecommand \@@href[1]{\endgroup#1\@@endlink}%
\providecommand \@sanitize@url [0]{\catcode `\\12\catcode `\$12\catcode
  `\&12\catcode `\#12\catcode `\^12\catcode `\_12\catcode `\%12\relax}%
\providecommand \@@startlink[1]{}%
\providecommand \@@endlink[0]{}%
\providecommand \url  [0]{\begingroup\@sanitize@url \@url }%
\providecommand \@url [1]{\endgroup\@href {#1}{\urlprefix }}%
\providecommand \urlprefix  [0]{URL }%
\providecommand \Eprint [0]{\href }%
\providecommand \doibase [0]{http://dx.doi.org/}%
\providecommand \selectlanguage [0]{\@gobble}%
\providecommand \bibinfo  [0]{\@secondoftwo}%
\providecommand \bibfield  [0]{\@secondoftwo}%
\providecommand \translation [1]{[#1]}%
\providecommand \BibitemOpen [0]{}%
\providecommand \bibitemStop [0]{}%
\providecommand \bibitemNoStop [0]{.\EOS\space}%
\providecommand \EOS [0]{\spacefactor3000\relax}%
\providecommand \BibitemShut  [1]{\csname bibitem#1\endcsname}%
\let\auto@bib@innerbib\@empty
\bibitem [{\citenamefont {Kamihara}\ \emph {et~al.}(2008)\citenamefont
  {Kamihara}, \citenamefont {Watanabe}, \citenamefont {Hirano},\ and\
  \citenamefont {Hosono}}]{Kamihara08JACS}%
  \BibitemOpen
  \bibfield  {author} {\bibinfo {author} {\bibfnamefont {Y.}~\bibnamefont
  {Kamihara}}, \bibinfo {author} {\bibfnamefont {T.}~\bibnamefont {Watanabe}},
  \bibinfo {author} {\bibfnamefont {M.}~\bibnamefont {Hirano}}, \ and\ \bibinfo
  {author} {\bibfnamefont {H.}~\bibnamefont {Hosono}},\ }\href {\doibase
  10.1021/ja800073m} {\bibfield  {journal} {\bibinfo  {journal} {J. Am. Chem.
  Soc.}\ }\textbf {\bibinfo {volume} {130}},\ \bibinfo {pages} {3296} (\bibinfo
  {year} {2008})}\BibitemShut {NoStop}%
\bibitem [{\citenamefont {Hirschfeld}\ \emph {et~al.}(2011)\citenamefont
  {Hirschfeld}, \citenamefont {Korshunov},\ and\ \citenamefont
  {Mazin}}]{Hirschfeld11RPP}%
  \BibitemOpen
  \bibfield  {author} {\bibinfo {author} {\bibfnamefont {P.~J.}\ \bibnamefont
  {Hirschfeld}}, \bibinfo {author} {\bibfnamefont {M.~M.}\ \bibnamefont
  {Korshunov}}, \ and\ \bibinfo {author} {\bibfnamefont {I.~I.}\ \bibnamefont
  {Mazin}},\ }\href {http://stacks.iop.org/0034-4885/74/i=12/a=124508}
  {\bibfield  {journal} {\bibinfo  {journal} {Rep. Prog. Phys.}\ }\textbf
  {\bibinfo {volume} {74}},\ \bibinfo {pages} {124508} (\bibinfo {year}
  {2011})}\BibitemShut {NoStop}%
\bibitem [{\citenamefont {Hashimoto}\ \emph {et~al.}(2009)\citenamefont
  {Hashimoto}, \citenamefont {Shibauchi}, \citenamefont {Kasahara},
  \citenamefont {Ikada}, \citenamefont {Tonegawa}, \citenamefont {Kato},
  \citenamefont {Okazaki}, \citenamefont {van~der Beek}, \citenamefont
  {Konczykowski}, \citenamefont {Takeya}, \citenamefont {Hirata}, \citenamefont
  {Terashima},\ and\ \citenamefont {Matsuda}}]{Hashimoto09PRL}%
  \BibitemOpen
  \bibfield  {author} {\bibinfo {author} {\bibfnamefont {K.}~\bibnamefont
  {Hashimoto}}, \bibinfo {author} {\bibfnamefont {T.}~\bibnamefont
  {Shibauchi}}, \bibinfo {author} {\bibfnamefont {S.}~\bibnamefont {Kasahara}},
  \bibinfo {author} {\bibfnamefont {K.}~\bibnamefont {Ikada}}, \bibinfo
  {author} {\bibfnamefont {S.}~\bibnamefont {Tonegawa}}, \bibinfo {author}
  {\bibfnamefont {T.}~\bibnamefont {Kato}}, \bibinfo {author} {\bibfnamefont
  {R.}~\bibnamefont {Okazaki}}, \bibinfo {author} {\bibfnamefont {C.~J.}\
  \bibnamefont {van~der Beek}}, \bibinfo {author} {\bibfnamefont
  {M.}~\bibnamefont {Konczykowski}}, \bibinfo {author} {\bibfnamefont
  {H.}~\bibnamefont {Takeya}}, \bibinfo {author} {\bibfnamefont
  {K.}~\bibnamefont {Hirata}}, \bibinfo {author} {\bibfnamefont
  {T.}~\bibnamefont {Terashima}}, \ and\ \bibinfo {author} {\bibfnamefont
  {Y.}~\bibnamefont {Matsuda}},\ }\href {\doibase
  10.1103/PhysRevLett.102.207001} {\bibfield  {journal} {\bibinfo  {journal}
  {Phys. Rev. Lett.}\ }\textbf {\bibinfo {volume} {102}},\ \bibinfo {pages}
  {207001} (\bibinfo {year} {2009})}\BibitemShut {NoStop}%
\bibitem [{\citenamefont {Mu}\ \emph {et~al.}(2009)\citenamefont {Mu},
  \citenamefont {Luo}, \citenamefont {Wang}, \citenamefont {Shan},
  \citenamefont {Ren},\ and\ \citenamefont {Wen}}]{Mu09PRB}%
  \BibitemOpen
  \bibfield  {author} {\bibinfo {author} {\bibfnamefont {G.}~\bibnamefont
  {Mu}}, \bibinfo {author} {\bibfnamefont {H.}~\bibnamefont {Luo}}, \bibinfo
  {author} {\bibfnamefont {Z.}~\bibnamefont {Wang}}, \bibinfo {author}
  {\bibfnamefont {L.}~\bibnamefont {Shan}}, \bibinfo {author} {\bibfnamefont
  {C.}~\bibnamefont {Ren}}, \ and\ \bibinfo {author} {\bibfnamefont {H.-H.}\
  \bibnamefont {Wen}},\ }\href {\doibase 10.1103/PhysRevB.79.174501} {\bibfield
   {journal} {\bibinfo  {journal} {Phys. Rev. B}\ }\textbf {\bibinfo {volume}
  {79}},\ \bibinfo {pages} {174501} (\bibinfo {year} {2009})}\BibitemShut
  {NoStop}%
\bibitem [{\citenamefont {Popovich}\ \emph {et~al.}(2010)\citenamefont
  {Popovich}, \citenamefont {Boris}, \citenamefont {Dolgov}, \citenamefont
  {Golubov}, \citenamefont {Sun}, \citenamefont {Lin}, \citenamefont {Kremer},\
  and\ \citenamefont {Keimer}}]{Popovich10PRL}%
  \BibitemOpen
  \bibfield  {author} {\bibinfo {author} {\bibfnamefont {P.}~\bibnamefont
  {Popovich}}, \bibinfo {author} {\bibfnamefont {A.~V.}\ \bibnamefont {Boris}},
  \bibinfo {author} {\bibfnamefont {O.~V.}\ \bibnamefont {Dolgov}}, \bibinfo
  {author} {\bibfnamefont {A.~A.}\ \bibnamefont {Golubov}}, \bibinfo {author}
  {\bibfnamefont {D.~L.}\ \bibnamefont {Sun}}, \bibinfo {author} {\bibfnamefont
  {C.~T.}\ \bibnamefont {Lin}}, \bibinfo {author} {\bibfnamefont {R.~K.}\
  \bibnamefont {Kremer}}, \ and\ \bibinfo {author} {\bibfnamefont
  {B.}~\bibnamefont {Keimer}},\ }\href {\doibase
  10.1103/PhysRevLett.105.027003} {\bibfield  {journal} {\bibinfo  {journal}
  {Phys. Rev. Lett.}\ }\textbf {\bibinfo {volume} {105}},\ \bibinfo {pages}
  {027003} (\bibinfo {year} {2010})}\BibitemShut {NoStop}%
\bibitem [{\citenamefont {Luo}\ \emph {et~al.}(2009)\citenamefont {Luo},
  \citenamefont {Tanatar}, \citenamefont {Reid}, \citenamefont {Shakeripour},
  \citenamefont {Doiron-Leyraud}, \citenamefont {Ni}, \citenamefont {Bud'ko},
  \citenamefont {Canfield}, \citenamefont {Luo}, \citenamefont {Wang},
  \citenamefont {Wen}, \citenamefont {Prozorov},\ and\ \citenamefont
  {Taillefer}}]{Luo09PRB}%
  \BibitemOpen
  \bibfield  {author} {\bibinfo {author} {\bibfnamefont {X.~G.}\ \bibnamefont
  {Luo}}, \bibinfo {author} {\bibfnamefont {M.~A.}\ \bibnamefont {Tanatar}},
  \bibinfo {author} {\bibfnamefont {J.-P.}\ \bibnamefont {Reid}}, \bibinfo
  {author} {\bibfnamefont {H.}~\bibnamefont {Shakeripour}}, \bibinfo {author}
  {\bibfnamefont {N.}~\bibnamefont {Doiron-Leyraud}}, \bibinfo {author}
  {\bibfnamefont {N.}~\bibnamefont {Ni}}, \bibinfo {author} {\bibfnamefont
  {S.~L.}\ \bibnamefont {Bud'ko}}, \bibinfo {author} {\bibfnamefont {P.~C.}\
  \bibnamefont {Canfield}}, \bibinfo {author} {\bibfnamefont {H.}~\bibnamefont
  {Luo}}, \bibinfo {author} {\bibfnamefont {Z.}~\bibnamefont {Wang}}, \bibinfo
  {author} {\bibfnamefont {H.-H.}\ \bibnamefont {Wen}}, \bibinfo {author}
  {\bibfnamefont {R.}~\bibnamefont {Prozorov}}, \ and\ \bibinfo {author}
  {\bibfnamefont {L.}~\bibnamefont {Taillefer}},\ }\href {\doibase
  10.1103/PhysRevB.80.140503} {\bibfield  {journal} {\bibinfo  {journal} {Phys.
  Rev. B}\ }\textbf {\bibinfo {volume} {80}},\ \bibinfo {pages} {140503}
  (\bibinfo {year} {2009})}\BibitemShut {NoStop}%
\bibitem [{\citenamefont {Yashima}\ \emph {et~al.}(2009)\citenamefont
  {Yashima}, \citenamefont {Nishimura}, \citenamefont {Mukuda}, \citenamefont
  {Kitaoka}, \citenamefont {Miyazawa}, \citenamefont {Shirage}, \citenamefont
  {Kihou}, \citenamefont {Kito}, \citenamefont {Eisaki},\ and\ \citenamefont
  {Iyo}}]{Yashima09JPSJ}%
  \BibitemOpen
  \bibfield  {author} {\bibinfo {author} {\bibfnamefont {M.}~\bibnamefont
  {Yashima}}, \bibinfo {author} {\bibfnamefont {H.}~\bibnamefont {Nishimura}},
  \bibinfo {author} {\bibfnamefont {H.}~\bibnamefont {Mukuda}}, \bibinfo
  {author} {\bibfnamefont {Y.}~\bibnamefont {Kitaoka}}, \bibinfo {author}
  {\bibfnamefont {K.}~\bibnamefont {Miyazawa}}, \bibinfo {author}
  {\bibfnamefont {P.~M.}\ \bibnamefont {Shirage}}, \bibinfo {author}
  {\bibfnamefont {K.}~\bibnamefont {Kihou}}, \bibinfo {author} {\bibfnamefont
  {H.}~\bibnamefont {Kito}}, \bibinfo {author} {\bibfnamefont {H.}~\bibnamefont
  {Eisaki}}, \ and\ \bibinfo {author} {\bibfnamefont {A.}~\bibnamefont {Iyo}},\
  }\href {\doibase 10.1143/JPSJ.78.103702} {\bibfield  {journal} {\bibinfo
  {journal} {J. Phys. Soc. Jpn.}\ }\textbf {\bibinfo {volume} {78}},\ \bibinfo
  {pages} {103702} (\bibinfo {year} {2009})}\BibitemShut {NoStop}%
\bibitem [{\citenamefont {Fukazawa}\ \emph {et~al.}(2009)\citenamefont
  {Fukazawa}, \citenamefont {Yamada}, \citenamefont {Kondo}, \citenamefont
  {Saito}, \citenamefont {Kohori}, \citenamefont {Kuga}, \citenamefont
  {Matsumoto}, \citenamefont {Nakatsuji}, \citenamefont {Kito}, \citenamefont
  {Shirage}, \citenamefont {Kihou}, \citenamefont {Takeshita}, \citenamefont
  {Lee}, \citenamefont {Iyo},\ and\ \citenamefont
  {Eisaki}}]{Fukazawa09JPSJ_KFA}%
  \BibitemOpen
  \bibfield  {author} {\bibinfo {author} {\bibfnamefont {H.}~\bibnamefont
  {Fukazawa}}, \bibinfo {author} {\bibfnamefont {Y.}~\bibnamefont {Yamada}},
  \bibinfo {author} {\bibfnamefont {K.}~\bibnamefont {Kondo}}, \bibinfo
  {author} {\bibfnamefont {T.}~\bibnamefont {Saito}}, \bibinfo {author}
  {\bibfnamefont {Y.}~\bibnamefont {Kohori}}, \bibinfo {author} {\bibfnamefont
  {K.}~\bibnamefont {Kuga}}, \bibinfo {author} {\bibfnamefont {Y.}~\bibnamefont
  {Matsumoto}}, \bibinfo {author} {\bibfnamefont {S.}~\bibnamefont
  {Nakatsuji}}, \bibinfo {author} {\bibfnamefont {H.}~\bibnamefont {Kito}},
  \bibinfo {author} {\bibfnamefont {P.~M.}\ \bibnamefont {Shirage}}, \bibinfo
  {author} {\bibfnamefont {K.}~\bibnamefont {Kihou}}, \bibinfo {author}
  {\bibfnamefont {N.}~\bibnamefont {Takeshita}}, \bibinfo {author}
  {\bibfnamefont {C.~H.}\ \bibnamefont {Lee}}, \bibinfo {author} {\bibfnamefont
  {A.}~\bibnamefont {Iyo}}, \ and\ \bibinfo {author} {\bibfnamefont
  {H.}~\bibnamefont {Eisaki}},\ }\href {\doibase 10.1143/JPSJ.78.083712}
  {\bibfield  {journal} {\bibinfo  {journal} {J. Phys. Soc. Jpn.}\ }\textbf
  {\bibinfo {volume} {78}},\ \bibinfo {pages} {083712} (\bibinfo {year}
  {2009})}\BibitemShut {NoStop}%
\bibitem [{\citenamefont {Hashimoto}\ \emph {et~al.}(2010)\citenamefont
  {Hashimoto}, \citenamefont {Serafin}, \citenamefont {Tonegawa}, \citenamefont
  {Katsumata}, \citenamefont {Okazaki}, \citenamefont {Saito}, \citenamefont
  {Fukazawa}, \citenamefont {Kohori}, \citenamefont {Kihou}, \citenamefont
  {Lee}, \citenamefont {Iyo}, \citenamefont {Eisaki}, \citenamefont {Ikeda},
  \citenamefont {Matsuda}, \citenamefont {Carrington},\ and\ \citenamefont
  {Shibauchi}}]{Hashimoto10PRB}%
  \BibitemOpen
  \bibfield  {author} {\bibinfo {author} {\bibfnamefont {K.}~\bibnamefont
  {Hashimoto}}, \bibinfo {author} {\bibfnamefont {A.}~\bibnamefont {Serafin}},
  \bibinfo {author} {\bibfnamefont {S.}~\bibnamefont {Tonegawa}}, \bibinfo
  {author} {\bibfnamefont {R.}~\bibnamefont {Katsumata}}, \bibinfo {author}
  {\bibfnamefont {R.}~\bibnamefont {Okazaki}}, \bibinfo {author} {\bibfnamefont
  {T.}~\bibnamefont {Saito}}, \bibinfo {author} {\bibfnamefont
  {H.}~\bibnamefont {Fukazawa}}, \bibinfo {author} {\bibfnamefont
  {Y.}~\bibnamefont {Kohori}}, \bibinfo {author} {\bibfnamefont
  {K.}~\bibnamefont {Kihou}}, \bibinfo {author} {\bibfnamefont {C.~H.}\
  \bibnamefont {Lee}}, \bibinfo {author} {\bibfnamefont {A.}~\bibnamefont
  {Iyo}}, \bibinfo {author} {\bibfnamefont {H.}~\bibnamefont {Eisaki}},
  \bibinfo {author} {\bibfnamefont {H.}~\bibnamefont {Ikeda}}, \bibinfo
  {author} {\bibfnamefont {Y.}~\bibnamefont {Matsuda}}, \bibinfo {author}
  {\bibfnamefont {A.}~\bibnamefont {Carrington}}, \ and\ \bibinfo {author}
  {\bibfnamefont {T.}~\bibnamefont {Shibauchi}},\ }\href {\doibase
  10.1103/PhysRevB.82.014526} {\bibfield  {journal} {\bibinfo  {journal} {Phys.
  Rev. B}\ }\textbf {\bibinfo {volume} {82}},\ \bibinfo {pages} {014526}
  (\bibinfo {year} {2010})}\BibitemShut {NoStop}%
\bibitem [{\citenamefont {Dong}\ \emph {et~al.}(2010)\citenamefont {Dong},
  \citenamefont {Zhou}, \citenamefont {Guan}, \citenamefont {Zhang},
  \citenamefont {Dai}, \citenamefont {Qiu}, \citenamefont {Wang}, \citenamefont
  {He}, \citenamefont {Chen},\ and\ \citenamefont {Li}}]{Dong10PRL}%
  \BibitemOpen
  \bibfield  {author} {\bibinfo {author} {\bibfnamefont {J.~K.}\ \bibnamefont
  {Dong}}, \bibinfo {author} {\bibfnamefont {S.~Y.}\ \bibnamefont {Zhou}},
  \bibinfo {author} {\bibfnamefont {T.~Y.}\ \bibnamefont {Guan}}, \bibinfo
  {author} {\bibfnamefont {H.}~\bibnamefont {Zhang}}, \bibinfo {author}
  {\bibfnamefont {Y.~F.}\ \bibnamefont {Dai}}, \bibinfo {author} {\bibfnamefont
  {X.}~\bibnamefont {Qiu}}, \bibinfo {author} {\bibfnamefont {X.~F.}\
  \bibnamefont {Wang}}, \bibinfo {author} {\bibfnamefont {Y.}~\bibnamefont
  {He}}, \bibinfo {author} {\bibfnamefont {X.~H.}\ \bibnamefont {Chen}}, \ and\
  \bibinfo {author} {\bibfnamefont {S.~Y.}\ \bibnamefont {Li}},\ }\href
  {\doibase 10.1103/PhysRevLett.104.087005} {\bibfield  {journal} {\bibinfo
  {journal} {Phys. Rev. Lett.}\ }\textbf {\bibinfo {volume} {104}},\ \bibinfo
  {pages} {087005} (\bibinfo {year} {2010})}\BibitemShut {NoStop}%
\bibitem [{\citenamefont {Reid}\ \emph {et~al.}(2012)\citenamefont {Reid},
  \citenamefont {Tanatar}, \citenamefont {Juneau-Fecteau}, \citenamefont
  {Gordon}, \citenamefont {de~Cotret}, \citenamefont {Doiron-Leyraud},
  \citenamefont {Saito}, \citenamefont {Fukazawa}, \citenamefont {Kohori},
  \citenamefont {Kihou}, \citenamefont {Lee}, \citenamefont {Iyo},
  \citenamefont {Eisaki}, \citenamefont {Prozorov},\ and\ \citenamefont
  {Taillefer}}]{Reid12PRL}%
  \BibitemOpen
  \bibfield  {author} {\bibinfo {author} {\bibfnamefont {J.-P.}\ \bibnamefont
  {Reid}}, \bibinfo {author} {\bibfnamefont {M.~A.}\ \bibnamefont {Tanatar}},
  \bibinfo {author} {\bibfnamefont {A.}~\bibnamefont {Juneau-Fecteau}},
  \bibinfo {author} {\bibfnamefont {R.~T.}\ \bibnamefont {Gordon}}, \bibinfo
  {author} {\bibfnamefont {S.~R.}\ \bibnamefont {de~Cotret}}, \bibinfo {author}
  {\bibfnamefont {N.}~\bibnamefont {Doiron-Leyraud}}, \bibinfo {author}
  {\bibfnamefont {T.}~\bibnamefont {Saito}}, \bibinfo {author} {\bibfnamefont
  {H.}~\bibnamefont {Fukazawa}}, \bibinfo {author} {\bibfnamefont
  {Y.}~\bibnamefont {Kohori}}, \bibinfo {author} {\bibfnamefont
  {K.}~\bibnamefont {Kihou}}, \bibinfo {author} {\bibfnamefont {C.~H.}\
  \bibnamefont {Lee}}, \bibinfo {author} {\bibfnamefont {A.}~\bibnamefont
  {Iyo}}, \bibinfo {author} {\bibfnamefont {H.}~\bibnamefont {Eisaki}},
  \bibinfo {author} {\bibfnamefont {R.}~\bibnamefont {Prozorov}}, \ and\
  \bibinfo {author} {\bibfnamefont {L.}~\bibnamefont {Taillefer}},\ }\href
  {\doibase 10.1103/PhysRevLett.109.087001} {\bibfield  {journal} {\bibinfo
  {journal} {Phys. Rev. Lett.}\ }\textbf {\bibinfo {volume} {109}},\ \bibinfo
  {pages} {087001} (\bibinfo {year} {2012})}\BibitemShut {NoStop}%
\bibitem [{\citenamefont {Okazaki}\ \emph {et~al.}(2012)\citenamefont
  {Okazaki}, \citenamefont {Ota}, \citenamefont {Kotani}, \citenamefont
  {Malaeb}, \citenamefont {Ishida}, \citenamefont {Shimojima}, \citenamefont
  {Kiss}, \citenamefont {Watanabe}, \citenamefont {Chen}, \citenamefont
  {Kihou}, \citenamefont {Lee}, \citenamefont {Iyo}, \citenamefont {Eisaki},
  \citenamefont {Saito}, \citenamefont {Fukazawa}, \citenamefont {Kohori},
  \citenamefont {Hashimoto}, \citenamefont {Shibauchi}, \citenamefont
  {Matsuda}, \citenamefont {Ikeda}, \citenamefont {Miyahara}, \citenamefont
  {Arita}, \citenamefont {Chainani},\ and\ \citenamefont
  {Shin}}]{Okazaki12Science}%
  \BibitemOpen
  \bibfield  {author} {\bibinfo {author} {\bibfnamefont {K.}~\bibnamefont
  {Okazaki}}, \bibinfo {author} {\bibfnamefont {Y.}~\bibnamefont {Ota}},
  \bibinfo {author} {\bibfnamefont {Y.}~\bibnamefont {Kotani}}, \bibinfo
  {author} {\bibfnamefont {W.}~\bibnamefont {Malaeb}}, \bibinfo {author}
  {\bibfnamefont {Y.}~\bibnamefont {Ishida}}, \bibinfo {author} {\bibfnamefont
  {T.}~\bibnamefont {Shimojima}}, \bibinfo {author} {\bibfnamefont
  {T.}~\bibnamefont {Kiss}}, \bibinfo {author} {\bibfnamefont {S.}~\bibnamefont
  {Watanabe}}, \bibinfo {author} {\bibfnamefont {C.-T.}\ \bibnamefont {Chen}},
  \bibinfo {author} {\bibfnamefont {K.}~\bibnamefont {Kihou}}, \bibinfo
  {author} {\bibfnamefont {C.~H.}\ \bibnamefont {Lee}}, \bibinfo {author}
  {\bibfnamefont {A.}~\bibnamefont {Iyo}}, \bibinfo {author} {\bibfnamefont
  {H.}~\bibnamefont {Eisaki}}, \bibinfo {author} {\bibfnamefont
  {T.}~\bibnamefont {Saito}}, \bibinfo {author} {\bibfnamefont
  {H.}~\bibnamefont {Fukazawa}}, \bibinfo {author} {\bibfnamefont
  {Y.}~\bibnamefont {Kohori}}, \bibinfo {author} {\bibfnamefont
  {K.}~\bibnamefont {Hashimoto}}, \bibinfo {author} {\bibfnamefont
  {T.}~\bibnamefont {Shibauchi}}, \bibinfo {author} {\bibfnamefont
  {Y.}~\bibnamefont {Matsuda}}, \bibinfo {author} {\bibfnamefont
  {H.}~\bibnamefont {Ikeda}}, \bibinfo {author} {\bibfnamefont
  {H.}~\bibnamefont {Miyahara}}, \bibinfo {author} {\bibfnamefont
  {R.}~\bibnamefont {Arita}}, \bibinfo {author} {\bibfnamefont
  {A.}~\bibnamefont {Chainani}}, \ and\ \bibinfo {author} {\bibfnamefont
  {S.}~\bibnamefont {Shin}},\ }\href {\doibase 10.1126/science.1222793}
  {\bibfield  {journal} {\bibinfo  {journal} {Science}\ }\textbf {\bibinfo
  {volume} {337}},\ \bibinfo {pages} {1314} (\bibinfo {year}
  {2012})}\BibitemShut {NoStop}%
\bibitem [{\citenamefont {Kittaka}\ \emph {et~al.}(2014)\citenamefont
  {Kittaka}, \citenamefont {Aoki}, \citenamefont {Kase}, \citenamefont
  {Sakakibara}, \citenamefont {Saito}, \citenamefont {Fukazawa}, \citenamefont
  {Kohori}, \citenamefont {Kihou}, \citenamefont {Lee}, \citenamefont {Iyo},
  \citenamefont {Eisaki}, \citenamefont {Deguchi}, \citenamefont {Sato},
  \citenamefont {Tsutsumi},\ and\ \citenamefont {Machida}}]{Kittaka14JPSJ}%
  \BibitemOpen
  \bibfield  {author} {\bibinfo {author} {\bibfnamefont {S.}~\bibnamefont
  {Kittaka}}, \bibinfo {author} {\bibfnamefont {Y.}~\bibnamefont {Aoki}},
  \bibinfo {author} {\bibfnamefont {N.}~\bibnamefont {Kase}}, \bibinfo {author}
  {\bibfnamefont {T.}~\bibnamefont {Sakakibara}}, \bibinfo {author}
  {\bibfnamefont {T.}~\bibnamefont {Saito}}, \bibinfo {author} {\bibfnamefont
  {H.}~\bibnamefont {Fukazawa}}, \bibinfo {author} {\bibfnamefont
  {Y.}~\bibnamefont {Kohori}}, \bibinfo {author} {\bibfnamefont
  {K.}~\bibnamefont {Kihou}}, \bibinfo {author} {\bibfnamefont {C.~H.}\
  \bibnamefont {Lee}}, \bibinfo {author} {\bibfnamefont {A.}~\bibnamefont
  {Iyo}}, \bibinfo {author} {\bibfnamefont {H.}~\bibnamefont {Eisaki}},
  \bibinfo {author} {\bibfnamefont {K.}~\bibnamefont {Deguchi}}, \bibinfo
  {author} {\bibfnamefont {N.~K.}\ \bibnamefont {Sato}}, \bibinfo {author}
  {\bibfnamefont {Y.}~\bibnamefont {Tsutsumi}}, \ and\ \bibinfo {author}
  {\bibfnamefont {K.}~\bibnamefont {Machida}},\ }\href {\doibase
  10.7566/JPSJ.83.013704} {\bibfield  {journal} {\bibinfo  {journal} {J. Phys.
  Soc. Jpn.}\ }\textbf {\bibinfo {volume} {83}},\ \bibinfo {pages} {013704}
  (\bibinfo {year} {2014})}\BibitemShut {NoStop}%
\bibitem [{\citenamefont {Thomale}\ \emph {et~al.}(2011)\citenamefont
  {Thomale}, \citenamefont {Platt}, \citenamefont {Hanke}, \citenamefont {Hu},\
  and\ \citenamefont {Bernevig}}]{Thomale11PRL}%
  \BibitemOpen
  \bibfield  {author} {\bibinfo {author} {\bibfnamefont {R.}~\bibnamefont
  {Thomale}}, \bibinfo {author} {\bibfnamefont {C.}~\bibnamefont {Platt}},
  \bibinfo {author} {\bibfnamefont {W.}~\bibnamefont {Hanke}}, \bibinfo
  {author} {\bibfnamefont {J.}~\bibnamefont {Hu}}, \ and\ \bibinfo {author}
  {\bibfnamefont {B.~A.}\ \bibnamefont {Bernevig}},\ }\href {\doibase
  10.1103/PhysRevLett.107.117001} {\bibfield  {journal} {\bibinfo  {journal}
  {Phys. Rev. Lett.}\ }\textbf {\bibinfo {volume} {107}},\ \bibinfo {pages}
  {117001} (\bibinfo {year} {2011})}\BibitemShut {NoStop}%
\bibitem [{\citenamefont {Maiti}\ \emph {et~al.}(2011)\citenamefont {Maiti},
  \citenamefont {Korshunov}, \citenamefont {Maier}, \citenamefont
  {Hirschfeld},\ and\ \citenamefont {Chubukov}}]{Maiti11PRL}%
  \BibitemOpen
  \bibfield  {author} {\bibinfo {author} {\bibfnamefont {S.}~\bibnamefont
  {Maiti}}, \bibinfo {author} {\bibfnamefont {M.~M.}\ \bibnamefont
  {Korshunov}}, \bibinfo {author} {\bibfnamefont {T.~A.}\ \bibnamefont
  {Maier}}, \bibinfo {author} {\bibfnamefont {P.~J.}\ \bibnamefont
  {Hirschfeld}}, \ and\ \bibinfo {author} {\bibfnamefont {A.~V.}\ \bibnamefont
  {Chubukov}},\ }\href {\doibase 10.1103/PhysRevLett.107.147002} {\bibfield
  {journal} {\bibinfo  {journal} {Phys. Rev. Lett.}\ }\textbf {\bibinfo
  {volume} {107}},\ \bibinfo {pages} {147002} (\bibinfo {year}
  {2011})}\BibitemShut {NoStop}%
\bibitem [{\citenamefont {Suzuki}\ \emph {et~al.}(2011)\citenamefont {Suzuki},
  \citenamefont {Usui},\ and\ \citenamefont {Kuroki}}]{Suzuki11PRB}%
  \BibitemOpen
  \bibfield  {author} {\bibinfo {author} {\bibfnamefont {K.}~\bibnamefont
  {Suzuki}}, \bibinfo {author} {\bibfnamefont {H.}~\bibnamefont {Usui}}, \ and\
  \bibinfo {author} {\bibfnamefont {K.}~\bibnamefont {Kuroki}},\ }\href
  {\doibase 10.1103/PhysRevB.84.144514} {\bibfield  {journal} {\bibinfo
  {journal} {Phys. Rev. B}\ }\textbf {\bibinfo {volume} {84}},\ \bibinfo
  {pages} {144514} (\bibinfo {year} {2011})}\BibitemShut {NoStop}%
\bibitem [{\citenamefont {Maiti}\ \emph {et~al.}(2012)\citenamefont {Maiti},
  \citenamefont {Korshunov},\ and\ \citenamefont {Chubukov}}]{Maiti12PRB}%
  \BibitemOpen
  \bibfield  {author} {\bibinfo {author} {\bibfnamefont {S.}~\bibnamefont
  {Maiti}}, \bibinfo {author} {\bibfnamefont {M.~M.}\ \bibnamefont
  {Korshunov}}, \ and\ \bibinfo {author} {\bibfnamefont {A.~V.}\ \bibnamefont
  {Chubukov}},\ }\href {\doibase 10.1103/PhysRevB.85.014511} {\bibfield
  {journal} {\bibinfo  {journal} {Phys. Rev. B}\ }\textbf {\bibinfo {volume}
  {85}},\ \bibinfo {pages} {014511} (\bibinfo {year} {2012})}\BibitemShut
  {NoStop}%
\bibitem [{\citenamefont {Tafti}\ \emph {et~al.}(2013)\citenamefont {Tafti},
  \citenamefont {Juneau-Fecteau}, \citenamefont {Delage}, \citenamefont
  {de~Cotret}, \citenamefont {Reid}, \citenamefont {Wang}, \citenamefont {Luo},
  \citenamefont {Chen}, \citenamefont {Doiron-Leyraud},\ and\ \citenamefont
  {Taillefer}}]{Tafti13NatPhys}%
  \BibitemOpen
  \bibfield  {author} {\bibinfo {author} {\bibfnamefont {F.~F.}\ \bibnamefont
  {Tafti}}, \bibinfo {author} {\bibfnamefont {A.}~\bibnamefont
  {Juneau-Fecteau}}, \bibinfo {author} {\bibfnamefont {M.-{\`E}.}\ \bibnamefont
  {Delage}}, \bibinfo {author} {\bibfnamefont {S.~R.}\ \bibnamefont
  {de~Cotret}}, \bibinfo {author} {\bibfnamefont {J.-P.}\ \bibnamefont {Reid}},
  \bibinfo {author} {\bibfnamefont {A.~F.}\ \bibnamefont {Wang}}, \bibinfo
  {author} {\bibfnamefont {X.-G.}\ \bibnamefont {Luo}}, \bibinfo {author}
  {\bibfnamefont {X.~H.}\ \bibnamefont {Chen}}, \bibinfo {author}
  {\bibfnamefont {N.}~\bibnamefont {Doiron-Leyraud}}, \ and\ \bibinfo {author}
  {\bibfnamefont {L.}~\bibnamefont {Taillefer}},\ }\href@noop {} {\bibfield
  {journal} {\bibinfo  {journal} {Nature Phys.}\ }\textbf {\bibinfo {volume}
  {9}},\ \bibinfo {pages} {349} (\bibinfo {year} {2013})}\BibitemShut {NoStop}%
\bibitem [{\citenamefont {Kihou}\ \emph {et~al.}(2010)\citenamefont {Kihou},
  \citenamefont {Saito}, \citenamefont {Ishida}, \citenamefont {Nakajima},
  \citenamefont {Tomioka}, \citenamefont {Fukazawa}, \citenamefont {Kohori},
  \citenamefont {Ito}, \citenamefont {Uchida}, \citenamefont {Iyo},
  \citenamefont {Lee},\ and\ \citenamefont {Eisaki}}]{Kihou10JPSJ}%
  \BibitemOpen
  \bibfield  {author} {\bibinfo {author} {\bibfnamefont {K.}~\bibnamefont
  {Kihou}}, \bibinfo {author} {\bibfnamefont {T.}~\bibnamefont {Saito}},
  \bibinfo {author} {\bibfnamefont {S.}~\bibnamefont {Ishida}}, \bibinfo
  {author} {\bibfnamefont {M.}~\bibnamefont {Nakajima}}, \bibinfo {author}
  {\bibfnamefont {Y.}~\bibnamefont {Tomioka}}, \bibinfo {author} {\bibfnamefont
  {H.}~\bibnamefont {Fukazawa}}, \bibinfo {author} {\bibfnamefont
  {Y.}~\bibnamefont {Kohori}}, \bibinfo {author} {\bibfnamefont
  {T.}~\bibnamefont {Ito}}, \bibinfo {author} {\bibfnamefont {S.}~\bibnamefont
  {Uchida}}, \bibinfo {author} {\bibfnamefont {A.}~\bibnamefont {Iyo}},
  \bibinfo {author} {\bibfnamefont {C.~H.}\ \bibnamefont {Lee}}, \ and\
  \bibinfo {author} {\bibfnamefont {H.}~\bibnamefont {Eisaki}},\ }\href
  {\doibase 10.1143/JPSJ.79.124713} {\bibfield  {journal} {\bibinfo  {journal}
  {J. Phys. Soc. Jpn.}\ }\textbf {\bibinfo {volume} {79}},\ \bibinfo {pages}
  {124713} (\bibinfo {year} {2010})}\BibitemShut {NoStop}%
\bibitem [{\citenamefont {Murata}\ \emph {et~al.}(2008)\citenamefont {Murata},
  \citenamefont {Yokogawa}, \citenamefont {Yoshino}, \citenamefont {Klotz},
  \citenamefont {Munsch}, \citenamefont {Irizawa}, \citenamefont {Nishiyama},
  \citenamefont {Iizuka}, \citenamefont {Nanba}, \citenamefont {Okada},
  \citenamefont {Shiraga},\ and\ \citenamefont {Aoyama}}]{Murata08RSI}%
  \BibitemOpen
  \bibfield  {author} {\bibinfo {author} {\bibfnamefont {K.}~\bibnamefont
  {Murata}}, \bibinfo {author} {\bibfnamefont {K.}~\bibnamefont {Yokogawa}},
  \bibinfo {author} {\bibfnamefont {H.}~\bibnamefont {Yoshino}}, \bibinfo
  {author} {\bibfnamefont {S.}~\bibnamefont {Klotz}}, \bibinfo {author}
  {\bibfnamefont {P.}~\bibnamefont {Munsch}}, \bibinfo {author} {\bibfnamefont
  {A.}~\bibnamefont {Irizawa}}, \bibinfo {author} {\bibfnamefont
  {M.}~\bibnamefont {Nishiyama}}, \bibinfo {author} {\bibfnamefont
  {K.}~\bibnamefont {Iizuka}}, \bibinfo {author} {\bibfnamefont
  {T.}~\bibnamefont {Nanba}}, \bibinfo {author} {\bibfnamefont
  {T.}~\bibnamefont {Okada}}, \bibinfo {author} {\bibfnamefont
  {Y.}~\bibnamefont {Shiraga}}, \ and\ \bibinfo {author} {\bibfnamefont
  {S.}~\bibnamefont {Aoyama}},\ }\href {\doibase 10.1063/1.2964117} {\bibfield
  {journal} {\bibinfo  {journal} {Rev. Sci. Instrum.}\ }\textbf {\bibinfo
  {volume} {79}},\ \bibinfo {eid} {085101} (\bibinfo {year}
  {2008})}\BibitemShut {NoStop}%
\bibitem [{\citenamefont {Terashima}\ \emph {et~al.}(2009)\citenamefont
  {Terashima}, \citenamefont {Tomita}, \citenamefont {Kimata}, \citenamefont
  {Satsukawa}, \citenamefont {Harada}, \citenamefont {Hazama}, \citenamefont
  {Uji}, \citenamefont {Suzuki}, \citenamefont {Matsumoto},\ and\ \citenamefont
  {Murata}}]{Terashima09JPSJ_EFA_erratum}%
  \BibitemOpen
  \bibfield  {author} {\bibinfo {author} {\bibfnamefont {T.}~\bibnamefont
  {Terashima}}, \bibinfo {author} {\bibfnamefont {M.}~\bibnamefont {Tomita}},
  \bibinfo {author} {\bibfnamefont {M.}~\bibnamefont {Kimata}}, \bibinfo
  {author} {\bibfnamefont {H.}~\bibnamefont {Satsukawa}}, \bibinfo {author}
  {\bibfnamefont {A.}~\bibnamefont {Harada}}, \bibinfo {author} {\bibfnamefont
  {K.}~\bibnamefont {Hazama}}, \bibinfo {author} {\bibfnamefont
  {S.}~\bibnamefont {Uji}}, \bibinfo {author} {\bibfnamefont {H.~S.}\
  \bibnamefont {Suzuki}}, \bibinfo {author} {\bibfnamefont {T.}~\bibnamefont
  {Matsumoto}}, \ and\ \bibinfo {author} {\bibfnamefont {K.}~\bibnamefont
  {Murata}},\ }\href {\doibase 10.1143/JPSJ.78.118001} {\bibfield  {journal}
  {\bibinfo  {journal} {J. Phys. Soc. Jpn.}\ }\textbf {\bibinfo {volume}
  {78}},\ \bibinfo {pages} {118001} (\bibinfo {year} {2009})}\BibitemShut
  {NoStop}%
\bibitem [{Note1()}]{Note1}%
  \BibitemOpen
  \bibinfo {note} {The ratio $B_{c2}/T_c$ would be related to the effective
  mass in the case of single-band superconductors without the Pauli
  paramagnetic effect, but its interpretation in multiband superconductors like
  KFe$_2$As$_2$ is not straightforward.}\BibitemShut {Stop}%
\bibitem [{\citenamefont {Taufour}\ \emph {et~al.}(2014)\citenamefont
  {Taufour}, \citenamefont {Foroozani}, \citenamefont {Lim}, \citenamefont
  {Tanatar}, \citenamefont {Kaluarachchi}, \citenamefont {Kim}, \citenamefont
  {Liu}, \citenamefont {Lograsso}, \citenamefont {Kogan}, \citenamefont
  {Prozorov}, \citenamefont {Budko}, \citenamefont {Schilling},\ and\
  \citenamefont {Canfield}}]{Taufour14condmat}%
  \BibitemOpen
  \bibfield  {author} {\bibinfo {author} {\bibfnamefont {V.}~\bibnamefont
  {Taufour}}, \bibinfo {author} {\bibfnamefont {N.}~\bibnamefont {Foroozani}},
  \bibinfo {author} {\bibfnamefont {J.}~\bibnamefont {Lim}}, \bibinfo {author}
  {\bibfnamefont {M.~A.}\ \bibnamefont {Tanatar}}, \bibinfo {author}
  {\bibfnamefont {U.}~\bibnamefont {Kaluarachchi}}, \bibinfo {author}
  {\bibfnamefont {S.~K.}\ \bibnamefont {Kim}}, \bibinfo {author} {\bibfnamefont
  {Y.}~\bibnamefont {Liu}}, \bibinfo {author} {\bibfnamefont {T.~A.}\
  \bibnamefont {Lograsso}}, \bibinfo {author} {\bibfnamefont {V.~G.}\
  \bibnamefont {Kogan}}, \bibinfo {author} {\bibfnamefont {R.}~\bibnamefont
  {Prozorov}}, \bibinfo {author} {\bibfnamefont {S.~L.}\ \bibnamefont {Budko}},
  \bibinfo {author} {\bibfnamefont {J.~S.}\ \bibnamefont {Schilling}}, \ and\
  \bibinfo {author} {\bibfnamefont {P.~C.}\ \bibnamefont {Canfield}},\
  }\href@noop {} {\bibfield  {journal} {\bibinfo  {journal} {arXiv:1402.7054}\
  } (\bibinfo {year} {2014})}\BibitemShut {NoStop}%
\bibitem [{\citenamefont {Bud'ko}\ \emph {et~al.}(2012)\citenamefont {Bud'ko},
  \citenamefont {Liu}, \citenamefont {Lograsso},\ and\ \citenamefont
  {Canfield}}]{Bud'ko12PRB}%
  \BibitemOpen
  \bibfield  {author} {\bibinfo {author} {\bibfnamefont {S.~L.}\ \bibnamefont
  {Bud'ko}}, \bibinfo {author} {\bibfnamefont {Y.}~\bibnamefont {Liu}},
  \bibinfo {author} {\bibfnamefont {T.~A.}\ \bibnamefont {Lograsso}}, \ and\
  \bibinfo {author} {\bibfnamefont {P.~C.}\ \bibnamefont {Canfield}},\ }\href
  {\doibase 10.1103/PhysRevB.86.224514} {\bibfield  {journal} {\bibinfo
  {journal} {Phys. Rev. B}\ }\textbf {\bibinfo {volume} {86}},\ \bibinfo
  {pages} {224514} (\bibinfo {year} {2012})}\BibitemShut {NoStop}%
\bibitem [{\citenamefont {Giamarchi}\ and\ \citenamefont
  {Bhattacharya}(2001)}]{Giamarchi01condmat}%
  \BibitemOpen
  \bibfield  {author} {\bibinfo {author} {\bibfnamefont {T.}~\bibnamefont
  {Giamarchi}}\ and\ \bibinfo {author} {\bibfnamefont {S.}~\bibnamefont
  {Bhattacharya}},\ }\href@noop {} {\bibfield  {journal} {\bibinfo  {journal}
  {arXiv:cond-mat/0111052}\ } (\bibinfo {year} {2001})}\BibitemShut {NoStop}%
\bibitem [{\citenamefont {Terashima}\ \emph {et~al.}(2013)\citenamefont
  {Terashima}, \citenamefont {Kurita}, \citenamefont {Kimata}, \citenamefont
  {Tomita}, \citenamefont {Tsuchiya}, \citenamefont {Imai}, \citenamefont
  {Sato}, \citenamefont {Kihou}, \citenamefont {Lee}, \citenamefont {Kito},
  \citenamefont {Eisaki}, \citenamefont {Iyo}, \citenamefont {Saito},
  \citenamefont {Fukazawa}, \citenamefont {Kohori}, \citenamefont {Harima},\
  and\ \citenamefont {Uji}}]{Terashima13PRBdHvA}%
  \BibitemOpen
  \bibfield  {author} {\bibinfo {author} {\bibfnamefont {T.}~\bibnamefont
  {Terashima}}, \bibinfo {author} {\bibfnamefont {N.}~\bibnamefont {Kurita}},
  \bibinfo {author} {\bibfnamefont {M.}~\bibnamefont {Kimata}}, \bibinfo
  {author} {\bibfnamefont {M.}~\bibnamefont {Tomita}}, \bibinfo {author}
  {\bibfnamefont {S.}~\bibnamefont {Tsuchiya}}, \bibinfo {author}
  {\bibfnamefont {M.}~\bibnamefont {Imai}}, \bibinfo {author} {\bibfnamefont
  {A.}~\bibnamefont {Sato}}, \bibinfo {author} {\bibfnamefont {K.}~\bibnamefont
  {Kihou}}, \bibinfo {author} {\bibfnamefont {C.~H.}\ \bibnamefont {Lee}},
  \bibinfo {author} {\bibfnamefont {H.}~\bibnamefont {Kito}}, \bibinfo {author}
  {\bibfnamefont {H.}~\bibnamefont {Eisaki}}, \bibinfo {author} {\bibfnamefont
  {A.}~\bibnamefont {Iyo}}, \bibinfo {author} {\bibfnamefont {T.}~\bibnamefont
  {Saito}}, \bibinfo {author} {\bibfnamefont {H.}~\bibnamefont {Fukazawa}},
  \bibinfo {author} {\bibfnamefont {Y.}~\bibnamefont {Kohori}}, \bibinfo
  {author} {\bibfnamefont {H.}~\bibnamefont {Harima}}, \ and\ \bibinfo {author}
  {\bibfnamefont {S.}~\bibnamefont {Uji}},\ }\href {\doibase
  10.1103/PhysRevB.87.224512} {\bibfield  {journal} {\bibinfo  {journal} {Phys.
  Rev. B}\ }\textbf {\bibinfo {volume} {87}},\ \bibinfo {pages} {224512}
  (\bibinfo {year} {2013})}\BibitemShut {NoStop}%
\bibitem [{\citenamefont {Sato}\ \emph {et~al.}(2009)\citenamefont {Sato},
  \citenamefont {Nakayama}, \citenamefont {Sekiba}, \citenamefont {Richard},
  \citenamefont {Xu}, \citenamefont {Souma}, \citenamefont {Takahashi},
  \citenamefont {Chen}, \citenamefont {Luo}, \citenamefont {Wang},\ and\
  \citenamefont {Ding}}]{Sato09PRL}%
  \BibitemOpen
  \bibfield  {author} {\bibinfo {author} {\bibfnamefont {T.}~\bibnamefont
  {Sato}}, \bibinfo {author} {\bibfnamefont {K.}~\bibnamefont {Nakayama}},
  \bibinfo {author} {\bibfnamefont {Y.}~\bibnamefont {Sekiba}}, \bibinfo
  {author} {\bibfnamefont {P.}~\bibnamefont {Richard}}, \bibinfo {author}
  {\bibfnamefont {Y.-M.}\ \bibnamefont {Xu}}, \bibinfo {author} {\bibfnamefont
  {S.}~\bibnamefont {Souma}}, \bibinfo {author} {\bibfnamefont
  {T.}~\bibnamefont {Takahashi}}, \bibinfo {author} {\bibfnamefont {G.~F.}\
  \bibnamefont {Chen}}, \bibinfo {author} {\bibfnamefont {J.~L.}\ \bibnamefont
  {Luo}}, \bibinfo {author} {\bibfnamefont {N.~L.}\ \bibnamefont {Wang}}, \
  and\ \bibinfo {author} {\bibfnamefont {H.}~\bibnamefont {Ding}},\ }\href
  {\doibase 10.1103/PhysRevLett.103.047002} {\bibfield  {journal} {\bibinfo
  {journal} {Phys. Rev. Lett.}\ }\textbf {\bibinfo {volume} {103}},\ \bibinfo
  {eid} {047002} (\bibinfo {year} {2009})}\BibitemShut {NoStop}%
\bibitem [{\citenamefont {Terashima}\ \emph {et~al.}(2010)\citenamefont
  {Terashima}, \citenamefont {Kimata}, \citenamefont {Kurita}, \citenamefont
  {Satsukawa}, \citenamefont {Harada}, \citenamefont {Hazama}, \citenamefont
  {Imai}, \citenamefont {Sato}, \citenamefont {Kihou}, \citenamefont {Lee},
  \citenamefont {Kito}, \citenamefont {Eisaki}, \citenamefont {Iyo},
  \citenamefont {Saito}, \citenamefont {Fukazawa}, \citenamefont {Kohori},
  \citenamefont {Harima},\ and\ \citenamefont {Uji}}]{Terashima10JPSJ}%
  \BibitemOpen
  \bibfield  {author} {\bibinfo {author} {\bibfnamefont {T.}~\bibnamefont
  {Terashima}}, \bibinfo {author} {\bibfnamefont {M.}~\bibnamefont {Kimata}},
  \bibinfo {author} {\bibfnamefont {N.}~\bibnamefont {Kurita}}, \bibinfo
  {author} {\bibfnamefont {H.}~\bibnamefont {Satsukawa}}, \bibinfo {author}
  {\bibfnamefont {A.}~\bibnamefont {Harada}}, \bibinfo {author} {\bibfnamefont
  {K.}~\bibnamefont {Hazama}}, \bibinfo {author} {\bibfnamefont
  {M.}~\bibnamefont {Imai}}, \bibinfo {author} {\bibfnamefont {A.}~\bibnamefont
  {Sato}}, \bibinfo {author} {\bibfnamefont {K.}~\bibnamefont {Kihou}},
  \bibinfo {author} {\bibfnamefont {C.~H.}\ \bibnamefont {Lee}}, \bibinfo
  {author} {\bibfnamefont {H.}~\bibnamefont {Kito}}, \bibinfo {author}
  {\bibfnamefont {H.}~\bibnamefont {Eisaki}}, \bibinfo {author} {\bibfnamefont
  {A.}~\bibnamefont {Iyo}}, \bibinfo {author} {\bibfnamefont {T.}~\bibnamefont
  {Saito}}, \bibinfo {author} {\bibfnamefont {H.}~\bibnamefont {Fukazawa}},
  \bibinfo {author} {\bibfnamefont {Y.}~\bibnamefont {Kohori}}, \bibinfo
  {author} {\bibfnamefont {H.}~\bibnamefont {Harima}}, \ and\ \bibinfo {author}
  {\bibfnamefont {S.}~\bibnamefont {Uji}},\ }\href {\doibase
  10.1143/JPSJ.79.053702} {\bibfield  {journal} {\bibinfo  {journal} {J. Phys.
  Soc. Jpn.}\ }\textbf {\bibinfo {volume} {79}},\ \bibinfo {pages} {053702}
  (\bibinfo {year} {2010})}\BibitemShut {NoStop}%
\bibitem [{\citenamefont {Yoshida}\ \emph {et~al.}(2011)\citenamefont
  {Yoshida}, \citenamefont {Nishi}, \citenamefont {Fujimori}, \citenamefont
  {Yi}, \citenamefont {Moore}, \citenamefont {Lu}, \citenamefont {Shen},
  \citenamefont {Kihou}, \citenamefont {Shirage}, \citenamefont {Kito},
  \citenamefont {Lee}, \citenamefont {Iyo}, \citenamefont {Eisaki},\ and\
  \citenamefont {Harima}}]{Yoshida11JPCS}%
  \BibitemOpen
  \bibfield  {author} {\bibinfo {author} {\bibfnamefont {T.}~\bibnamefont
  {Yoshida}}, \bibinfo {author} {\bibfnamefont {I.}~\bibnamefont {Nishi}},
  \bibinfo {author} {\bibfnamefont {A.}~\bibnamefont {Fujimori}}, \bibinfo
  {author} {\bibfnamefont {M.}~\bibnamefont {Yi}}, \bibinfo {author}
  {\bibfnamefont {R.~G.}\ \bibnamefont {Moore}}, \bibinfo {author}
  {\bibfnamefont {D.-H.}\ \bibnamefont {Lu}}, \bibinfo {author} {\bibfnamefont
  {Z.-X.}\ \bibnamefont {Shen}}, \bibinfo {author} {\bibfnamefont
  {K.}~\bibnamefont {Kihou}}, \bibinfo {author} {\bibfnamefont {P.~M.}\
  \bibnamefont {Shirage}}, \bibinfo {author} {\bibfnamefont {H.}~\bibnamefont
  {Kito}}, \bibinfo {author} {\bibfnamefont {C.~H.}\ \bibnamefont {Lee}},
  \bibinfo {author} {\bibfnamefont {A.}~\bibnamefont {Iyo}}, \bibinfo {author}
  {\bibfnamefont {H.}~\bibnamefont {Eisaki}}, \ and\ \bibinfo {author}
  {\bibfnamefont {H.}~\bibnamefont {Harima}},\ }\href {\doibase
  10.1016/j.jpcs.2010.10.064} {\bibfield  {journal} {\bibinfo  {journal} {J.
  Phys. Chem. Solids}\ }\textbf {\bibinfo {volume} {72}},\ \bibinfo {pages}
  {465 } (\bibinfo {year} {2011})}\BibitemShut {NoStop}%
\bibitem [{\citenamefont {Shoenberg}(1984)}]{Shoenberg84}%
  \BibitemOpen
  \bibfield  {author} {\bibinfo {author} {\bibfnamefont {D.}~\bibnamefont
  {Shoenberg}},\ }\href@noop {} {\emph {\bibinfo {title} {Magnetic oscillations
  in metals}}}\ (\bibinfo  {publisher} {Cambridge University Press},\ \bibinfo
  {address} {Cambridge},\ \bibinfo {year} {1984})\BibitemShut {NoStop}%
\bibitem [{\citenamefont {Kimber}\ \emph {et~al.}(2009)\citenamefont {Kimber},
  \citenamefont {Kreyssig}, \citenamefont {Zhang}, \citenamefont {Jeschke},
  \citenamefont {Valent\'{\i}}, \citenamefont {Yokaichiya}, \citenamefont
  {Colombier}, \citenamefont {Yan}, \citenamefont {Hansen}, \citenamefont
  {Chatterji}, \citenamefont {McQueeney}, \citenamefont {Canfield},
  \citenamefont {Goldman},\ and\ \citenamefont {Argyriou}}]{Kimber09NatMat}%
  \BibitemOpen
  \bibfield  {author} {\bibinfo {author} {\bibfnamefont {S.~A.~J.}\
  \bibnamefont {Kimber}}, \bibinfo {author} {\bibfnamefont {A.}~\bibnamefont
  {Kreyssig}}, \bibinfo {author} {\bibfnamefont {Y.-Z.}\ \bibnamefont {Zhang}},
  \bibinfo {author} {\bibfnamefont {H.~O.}\ \bibnamefont {Jeschke}}, \bibinfo
  {author} {\bibfnamefont {R.}~\bibnamefont {Valent\'{\i}}}, \bibinfo {author}
  {\bibfnamefont {F.}~\bibnamefont {Yokaichiya}}, \bibinfo {author}
  {\bibfnamefont {E.}~\bibnamefont {Colombier}}, \bibinfo {author}
  {\bibfnamefont {J.}~\bibnamefont {Yan}}, \bibinfo {author} {\bibfnamefont
  {T.~C.}\ \bibnamefont {Hansen}}, \bibinfo {author} {\bibfnamefont
  {T.}~\bibnamefont {Chatterji}}, \bibinfo {author} {\bibfnamefont {R.~J.}\
  \bibnamefont {McQueeney}}, \bibinfo {author} {\bibfnamefont {P.~C.}\
  \bibnamefont {Canfield}}, \bibinfo {author} {\bibfnamefont {A.~I.}\
  \bibnamefont {Goldman}}, \ and\ \bibinfo {author} {\bibfnamefont {D.~N.}\
  \bibnamefont {Argyriou}},\ }\href@noop {} {\bibfield  {journal} {\bibinfo
  {journal} {Nat. Mater.}\ }\textbf {\bibinfo {volume} {8}},\ \bibinfo {pages}
  {471} (\bibinfo {year} {2009})}\BibitemShut {NoStop}%
\bibitem [{\citenamefont {Nakamura}\ \emph {et~al.}(1996)\citenamefont
  {Nakamura}, \citenamefont {Moriya},\ and\ \citenamefont
  {Ueda}}]{Nakamura96JPSJ}%
  \BibitemOpen
  \bibfield  {author} {\bibinfo {author} {\bibfnamefont {S.}~\bibnamefont
  {Nakamura}}, \bibinfo {author} {\bibfnamefont {T.}~\bibnamefont {Moriya}}, \
  and\ \bibinfo {author} {\bibfnamefont {K.}~\bibnamefont {Ueda}},\ }\href
  {\doibase 10.1143/JPSJ.65.4026} {\bibfield  {journal} {\bibinfo  {journal}
  {J. Phys. Soc. Jpn.}\ }\textbf {\bibinfo {volume} {65}},\ \bibinfo {pages}
  {4026} (\bibinfo {year} {1996})}\BibitemShut {NoStop}%
\bibitem [{\citenamefont {Monthoux}\ and\ \citenamefont
  {Lonzarich}(2001)}]{Monthoux01PRB}%
  \BibitemOpen
  \bibfield  {author} {\bibinfo {author} {\bibfnamefont {P.}~\bibnamefont
  {Monthoux}}\ and\ \bibinfo {author} {\bibfnamefont {G.~G.}\ \bibnamefont
  {Lonzarich}},\ }\href {\doibase 10.1103/PhysRevB.63.054529} {\bibfield
  {journal} {\bibinfo  {journal} {Phys. Rev. B}\ }\textbf {\bibinfo {volume}
  {63}},\ \bibinfo {pages} {054529} (\bibinfo {year} {2001})}\BibitemShut
  {NoStop}%
\bibitem [{\citenamefont {Kuroki}\ and\ \citenamefont
  {Arita}(2001)}]{Kuroki01PRB}%
  \BibitemOpen
  \bibfield  {author} {\bibinfo {author} {\bibfnamefont {K.}~\bibnamefont
  {Kuroki}}\ and\ \bibinfo {author} {\bibfnamefont {R.}~\bibnamefont {Arita}},\
  }\href {\doibase 10.1103/PhysRevB.64.024501} {\bibfield  {journal} {\bibinfo
  {journal} {Phys. Rev. B}\ }\textbf {\bibinfo {volume} {64}},\ \bibinfo
  {pages} {024501} (\bibinfo {year} {2001})}\BibitemShut {NoStop}%
\end{thebibliography}
\end{document}